\documentclass[manuscript]{acmart}
\usepackage{graphicx}
\usepackage{algorithm}
\usepackage{algpseudocode}
\usepackage{multirow}
\usepackage{booktabs}
\usepackage{subcaption}

\AtBeginDocument{%
  }

\setcopyright{acmlicensed}
\copyrightyear{2018}
\acmYear{2018}
\acmDOI{XXXXXXX.XXXXXXX}

\acmConference[Conference acronym 'XX]{Make sure to enter the correct
  conference title from your rights confirmation emai}{June 03--05,
  2018}{Woodstock, NY}
\acmISBN{978-1-4503-XXXX-X/18/06}

\begin{document}

\title[Immersed in my Ideas]{Immersed in my Ideas: Using Virtual Reality and Multimodal Interactions to Visualize Users' Ideas and Thoughts}

\author{Yunhao Xing}
\email{xyunhao@nd.edu}
\affiliation{%
  \institution{University of Notre Dame}
  \city{Notre Dame}
  \state{Indiana}
  \country{USA}
  \postcode{46556}
}

\author{Jerrick Ban}
\email{jban@nd.edu}
\affiliation{%
  \institution{University of Notre Dame}
  \city{Notre Dame}
  \state{Indiana}
  \country{USA}
  \postcode{46556}
}

\author{Timothy D. Hubbard}
\email{thubbard@nd.edu}
\affiliation{%
  \institution{University of Notre Dame}
  \city{Notre Dame}
  \state{Indiana}
  \country{USA}
  \postcode{46556}
}

\author{Michael Villano}
\email{mvillan1@nd.edu}
\affiliation{%
  \institution{University of Notre Dame}
  \city{Notre Dame}
  \state{Indiana}
  \country{USA}
  \postcode{46556}
}

\author{Diego Gomez-Zara}
\email{dgomezara@nd.edu}
\affiliation{%
  \institution{University of Notre Dame}
  \city{Notre Dame}
  \state{Indiana}
  \country{USA}
  \postcode{46556}
}

\renewcommand{\shortauthors}{Xing et al.}

\begin{abstract}

This paper introduces VIVRA (Voice Interactive Virtual Reality Annotation), a VR application combining multimodal interaction with large language models (LLMs) to transform users' ideas into interactive 3D visualizations. VIVRA converts verbalized thoughts into "idea balloons" that summarize and expand on detected topics by an LLM. VIVRA allows users to verbalize their thoughts in real time or record their ideas to display the topics later. We evaluated the effectiveness of VIVRA in an exploratory study with 29 participants and a user study with 10 participants. Our results show that VIVRA enhanced users' ability to reflect on and develop ideas, achieving high levels of satisfaction, usability, and engagement. Participants valued VIVRA as a reflective tool for exploring personal thoughts and ideas. We discuss the potential advantages and uses of this application, highlighting the potential of combining immersive technologies with LLMs to create powerful ideation and reflection tools.

\end{abstract}

\ccsdesc[300]{Human-centered computing~Visualization systems and tools}
\ccsdesc[100]{Human-centered computing~Empirical studies in visualization}
\ccsdesc[500]{Human-centered computing~Human computer interaction (HCI)}
\ccsdesc[500]{Human-centered computing~Virtual reality}

\keywords{virtual reality, multimodal interaction, text visualization, large language model, reflection, creativity, idea generation}

\begin{teaserfigure}
  \includegraphics[width=\textwidth]{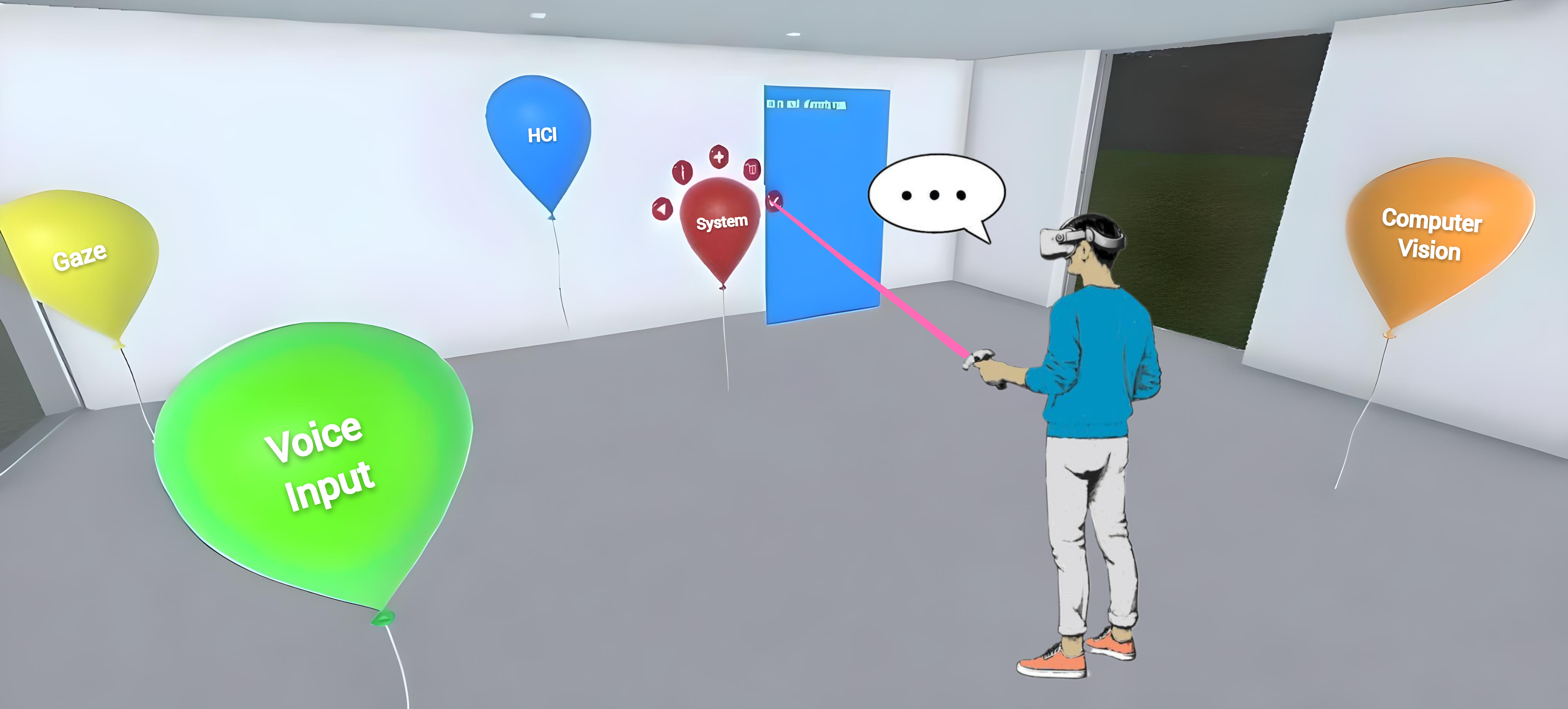}
  \caption{The main interface of VIVRA. Users can think aloud or play recordings in the environment. VIVRA will record, summarize, and visualize users' speech topics with the assistance of LLMs in the form of balloons. Users are able to interact with the environment with their voice, gaze, and controllers.}
  \Description{The main interface of VIVRA. Users can think aloud or play spoken recordings in the environment. VIVRA will record, transcribe, summarize, and visualize users' speech topics with the assistance of LLMs in the form of balloons. Users can move around and interact with the environment with their voice, gaze, and controllers.}
  \label{fig:teaser}
\end{teaserfigure}

\received{20 February 2007}
\received[revised]{12 March 2009}
\received[accepted]{5 June 2009}

\maketitle

\section{Introduction}
Analyzing and reflecting on our ideas can often be challenging tasks \cite{heong2012needs}. Typically, people require considerable effort to iterate over several ideas, opinions, thoughts, and facts \cite{adair2007art}. Moreover, materializing and organizing ideas demands finding connections among them, prioritizing and discarding certain ones, and producing new knowledge that extends them. Existing visualization techniques--such as mind maps, word clouds, or post-its--can help people synthesize and organize their ideas to obtain a better understanding \cite{Kucher2015,Hearst2023}. These techniques can facilitate the discovery, organization, combination, and discarding of ideas, enabling people to organize them better \cite{eppler2006comparison}. Despite their multiple advantages, these visualization techniques require considerable effort, learning, and skills to be helpful \cite{mladenovic2021impact,cao2016overview}. 

In this paper, we explore a novel approach that allows users to interact with their ideas using their voices and gaze in an immersive environment. Combining the capabilities of Virtual Reality (VR) and Large Language Models (LLMs), we develop a VR application called ``Voice Interactive Virtual Reality Annotation'' (VIVRA). This application allows users to visualize their ideas using multimodal interactions, including voice, gaze, and controllers. The goal of the system is to capture, summarize, expand, and represent users' main ideas as dynamic objects in a fully immersive 3D environment. Users of this system can think of their ideas aloud and see their ideas in real time. While users speak, the VIVRA system detects their ideas using a speech-recognition service and an LLM that summarizes their dialogues. Users can also create recordings and see emerging ideas in VR while listening to the recordings. To make the visualization dynamic and lucid, the current implementation displays each new idea as a balloon emerging from a portal on the floor. Users can interact with the balloons to create, delete, edit, and organize their ideas using their voice and the controllers. By using the LLM, VIVRA helps users link similar or contrasting ideas together. In designing VIVRA, our goal is to help users reflect on their ideas, which can help them understand, restructure, and critique their ideas on the fly \cite{schon1987educating}. 

The design of VIVRA embodies our vision of immersive and dynamic multimodal interactions between users and their ideas supported by LLMs and VR visualization techniques. First, multimodal interaction techniques, such as gaze and voice, can provide a convenient and engaging user experience. In designing VIVRA, we aim to integrate straightforward and intuitive instructions and multimodal operations that are easy to follow. Second, VIVRA should support a rich storytelling structure that can help users understand their ideas and apply changes to them. A well-designed VR system should enable linear narratives that display ideas in a sequence, as well as interactive narratives that allow users to create, navigate, and edit their ideas. Third, integrating visually engaging elements into the VIVRA system is crucial to fostering user engagement and usability \cite{jerald2015vr}. The visualizations employed in this application should also be informative and clear to facilitate a deeper understanding of users' ideas, ultimately supporting users' reflection and creativity. Lastly, VIVRA must ensure dynamic recordings, high-quality summaries, and effective visualizations to provide immediate feedback to its users. To achieve this goal, we employ LLMs to facilitate text data processing in real-time. By combining the immersive experience provided by VR and the LLMs' capabilities to analyze, summarize, and find new connections among concepts, VIVRA users can deepen their understanding and foster greater reflection on their ideas. 

To evaluate the effectiveness of VIVRA's approach, we conducted an exploratory study with 29 participants to evaluate its features and affordances. Our results from the post-treatment questionnaires and open-ended questions indicated that the participants enjoyed using an application such as VIVRA, and appreciated reflecting on their ideas while talking in real-time. To evaluate VIVRA's visualization features, we conducted a formal evaluation that tested different baseline visualizations (i.e., reading the original transcript and seeing a word cloud) with 10 participants. The participants first recorded their ideas and then listened to their recordings as VIVRA displayed the detected ideas emerging from the portals. The results were consistent with those of the first study, indicating that the participants scored VIVRA's visualization highly in usability, usefulness, and reflection activities.  

In summary, the contributions of this work are threefold: (a) a novel application to visualize and expand users' ideas and thoughts through a combination of speech detection, LLM processing, and 3D visualization in a VR environment; (b) a new approach to integrating multimodal interaction with voice, gaze, and controllers into idea visualization in VR; and (c) two empirical studies to evaluate the usability and effectiveness of VIVRA, including an exploratory analysis and a within-subjects user study that tested different visualizations.

\section{Related Work}
In this section, we review previous studies relevant to our research objectives. First, we discuss how previous research has explored user text input in VR/AR environments. We then describe several topic visualization approaches in interactive systems. Finally, we review how VR and LLM technologies can support users' reflections on their ideas.

\subsection{User Text Input in VR/AR Systems}
Inputting text in VR is challenging with conventional input devices since users cannot easily operate traditional devices designed for computers (e.g., keywords, mouse devices) while using a head-mounted device \cite{McGill2015, Mcgill2020}. To address these issues, many VR applications and operating systems provide virtual keyboards in which users use controllers or hand gestures to press the keys. However, these emulated keyboards are difficult to manipulate, and typing takes longer than normal keyboards \cite{Selection-based}. Previous work has developed multiple methods and techniques to improve text input within VR environments. Researchers have explored various alternative controller layouts for text input, such as circular \cite{PizzaText}, cubic \cite{Cubic}, and 12-key \cite{12-key} layouts, which diverge from the traditional QWERTY keyboard layout. In addition to virtual input methods, other innovative tools and solutions have integrated hardware devices. For example, ``Virtual Notepad'' \cite{VirtualNotepad} is a collection of VR interface tools that allow users to input text using a stylus pen in an immersive VR environment. In another study, Knierim et al. \cite{PhysicalKeyboards} developed a device that tracks the user's hand movements along with a physical keyboard. Another example is Kim et al.'s \cite{HoVR-Type} text entry method for integrating smartphones in VR environments. When a user's finger hovers on the smartphone screen, a transparent indicator shows up on the VR keyboard corresponding to their finger position. Lastly, Liu et al. \cite{PrinType} designed a technique that assigns different keys to different users' finger regions and utilizes a small thumb-worn sensor to recognize the keys.

Recent advances have expanded the use of wearable devices, such as smartwatches and smart glasses, to support text entry in VR. The ``PalmGesture'' system \cite{PalmGesture} uses an IR camera and a laser line projector to track users' drawing gestures on their palms, enabling them to handwrite phrases or draw symbols for predefined operations. Ahn et al. \cite{Smartwatch} employed a smartwatch as an indirect input device for AR smart glasses. Grossman et al. \cite{TypingOnGlasses} investigated the feasibility of a gesture-based text entry technique in smart eyewear, using the side touchpad as an input method. Lastly, ``DigiTap'' \cite{DigiTap} is a wrist-worn device that can sense thumb-to-finger taps, allowing robust eyes-free input. Although these hardware enhancements can significantly improve the speed and precision of user text entry in VR, they typically require additional efforts and resources to set up and employ. 


Voice-based text input techniques in VR systems have gained increasing attention. Bowman et al. \cite{bowman2002text} conducted a between-subjects study to compare the text input techniques in virtual reality. Participants had to type a word or phrase in VR using their voices, pens, tablets, and keyboards. The results showed that the voice input technique was the fastest. Another example is SWIFTER \cite{pick2016swifter}, a speech-based multimodal text entry system that allows users to enter text by voice as discrete segments and accumulate those into a larger text block. Moreover, recent VR devices such as Apple Vision Pro \cite{Apple} enable voice dictation features to help users enter text. This device also integrates gaze assistance to allow users to type using their voice in the place where they look. 

Inspired by previous work, we examine how multimodal inputs--including voice, controllers, and gaze--can enhance text entry in VR environments. Combining voice input with gaze-assisted could effectively address the limitations of relying solely on controller-based or gesture-based text input techniques.

\subsection{Topic Visualization}
Previous research has extensively worked on topic visualization techniques to facilitate the comprehension of ideas and text data. Visualizing keywords or topics enhances comprehension by presenting the text data in a clear and intuitive manner, making it a valuable tool in various domains such as NLP research, data analysis, and storytelling. ``Word clouds'' are one of the most common techniques for visualizing topics \cite{Heimerl2014}. They visually represent text data by displaying words in varying sizes according to their frequency, offering a clear summary of their content at a glance. For example, Cui et al. \cite{Cuiword} combined word cloud visualization and trend charts to illustrate the evolution of temporal content in a set of documents. Another example is the ``Concentri Cloud'' \cite{concentri}, a layered word cloud layout that could merge words from different text documents into a single visualization. Kaser et al. \cite{kaser2007tag} presented multiple word cloud visualization algorithms to improve the display of word clouds and show the relationship between different keywords, including tree-map structures, rectangle layouts, and bi-partitioning. ``ManiWordle'' \cite{koh2010maniwordle} added custom manipulation to word clouds so that users can have better control over the layout result. ``Prefix Tag Clouds'' \cite{burch2013prefix} is another example that includes relationships between words and transforms them into a subtree based on their prefix. Lastly, Xie et al. \cite{xie2023creating} created an animated version of a word cloud to give the words an emotional context and temporal relevance.

Another technique for presenting topics is the node-link structure \cite{saket2014node}, which displays relationships between entities by displaying them in a network-like structure. An example is ``Phrase net'' \cite{van2009mapping}, which displays a graph in which its nodes are words, and its edges are relationships defined by simple pattern matching or syntactic analysis. Another node-link structure is a ``mind map,'' which has numerous applications in teaching \cite{edwards2010mind, arulselvi2017mind, gagic2019implementation,abdel2017mind, dushkova2015use, loc2020using}, learning \cite{erdem2017mind,wickramasinghe2011effectiveness,willis2006mind,arulselvi2017mind}, and brainstorming \cite{abd2016brainstorming, alqasham2022effectiveness, budd2004mind}. Unlike traditional mind maps, Serra et al. \cite{serra2016behavioural} combined motion graphs to synthesize animation and mind maps as controllers to create new animation methods, where mind maps were used to decide when each motion should be displayed. Furthermore, VERITAS \cite{mindmapVR} is a VR application for reflective tasks in which users can use their controllers to create and interact with 3D mind maps. Lastly, Giraudeau et al. \cite{giraudeau2017towards} designed a mixed reality (MR) interface that allows users to create mind maps in MR and connect virtual nodes with physical objects.


In this work, we extend these previous techniques by introducing a novel way of topic visualization in the form of animated and immersive objects. We employ time and space to enhance the visualization of topics and keywords from users' ideas. The goal is to provide a more visually appealing and informative experience for users.

\subsection{Supporting reflection through VR and LLMs}
Schön introduced the concept of ``reflection-in-action," proposing that creativity can be nurtured by reflecting on one's actions during learning and practice activities \cite{schon1987educating}. This conceptualization highlights the importance of regularly analyzing experiences, identifying patterns, and gaining insight to support the creative process. As such, visualization can enable users to analyze, monitor, and adjust their practices over time. In particular, VR environments can offer users a way to engage with immersive scenarios that can be quickly adjusted. VR can promote spatial awareness and analysis, allowing users to engage with content and actions based on immersive experiences \cite{yang2018examining}. For example, Lee et al. \cite{lee2021using} developed a VR tool to support sketching in fashion design, and their study demonstrated that users were more creative using VR than using a 2D software application. Moreover, Yang et al. \cite{yang2018examining} found that creative prototypes developed in VR exhibited higher quality features compared to those created using pen-and-pencil methods. Taken together, immersive simulations and 3D digital objects can help users reflect on their ideas, information available, and outcomes.   

Using LLMs to support reflection has become increasingly popular due to their versatility and effectiveness in understanding, analyzing, and generating complex text data. The most common usage of LLMs for reflection is the inspiration and ideation process. Tholander et al. \cite{Design} organized a workshop and gave participants a design challenge with the support of AI tools to examine how LLMs can help in the design processes of ideation, early prototyping, and sketching. Their findings showed that LLMs could rapidly generate design alternatives. Another study conducted by Gero et al. \cite{gero2021sparks} investigated how LLMs can support science writing, which is an open-ended and constrained activity. Their results indicated that LLMs can serve different roles, including providing inspiration, translating sentences, and showing external perspectives. Furthermore, a recent study by Girotra et al. \cite{girotra2023ideas} found that OpenAI's ChatGPT-4 can generate ideas much faster and cheaper than students while maintaining, on average, higher quality. Another system is the ``Idea Machine'' \cite{di2022idea}, an LLM-based creativity support tool that expands, rewrites, or merges users' ideas to empower people to engage in idea-generation tasks. For group tasks, Gonzalez et al. \cite{gonzalez2024collaborative} developed a prototype tool that offers a shared canvas where group members and the LLM can share ideas in the form of virtual ``sticky notes.'' Yuan et al. \cite{Wordcraft} built a text editor based on an LLM that integrates various functions, including replacing, rewriting, elaborating, and filling. The results showed that the system made better and faster suggestions than the baselines and had a higher acceptance rate by users. Another example is ``BunCho'' \cite{BunCho}, a LLM-supported tool that helps with story co-creation in Japanese. This system utilizes LLM to help writers generate titles and synopses from keywords. They found that most writers enjoyed using the system and it broadened their stories by offering diverse suggestions. Furthermore, ``TaleBrush'' \cite{TaleBrush}, a story ideation tool that allows users to sketch a story arc by drawing a curve and use an LLM to generate a story based on the drawing. One last example is the ``CasualMapper'' \cite{CausalMapper}, which assists students in reasoning about the relationships between problems and solutions using casual maps. 

In this study, we aim to develop an application that fosters user reflection by enabling the visualization and expansion of their ideas and thoughts. Drawing inspiration from previous research, we have integrated LLMs with VR technology to create an immersive experience that allows users to interact with and learn from their ideas. This innovative approach offers a novel method for examining, understanding, and iterating thoughts and ideas through multimodal interactions.

\section{The VIVRA System}
\subsection{Design Goals}
We developed a VR application that provides a new visualization experience that helps users comprehend, summarize, organize, and expand their ideas and thoughts. The goal is to support reflection, enabling idea generation, critical thinking, and in-depth analysis of their ideas. Based on our analysis of the literature review, we identified four design goals that our system aims to achieve.

\paragraph{DG1: Integrating multimodal techniques to leverage user interaction with ideas and thoughts.} Various user interaction techniques in the virtual reality environment allow us to foster the interactions between users and digital content. Integrating multiple interaction data coming from users' bodies and movements, including gaze \cite{GazeSpeedup, GAVIN, Rajanna2018}, gestures \cite{PalmGesture, GazeSpeedup, Vulture, Apple}, and voice \cite{WalkingStick, Apple}, can leverage users' experience when viewing, creating, and editing their ideas. Using body interaction, such as voice and gaze, can facilitate an immersive experience to analyze topics and thoughts. Moreover, enabling multimodal input options allows users to conveniently examine the topics and ideas extracted from their transcripts by enabling multiple options to provide the input. It is crucial to avoid unnecessary input options that confuse users. Therefore, our first goal is to design simple and intuitive instructions and multimodal operations that users can easily follow.

\paragraph{DG2: Providing a rich narrative that seamlessly integrates both linear temporal and interactive real-time storytelling elements.} Users' ideas and thoughts can follow different kinds of narratives, which can be divided into linear or interactive \cite{schjerlund2018design}. While a linear narrative is a temporal sequence of events, an interactive narrative enables users to unfold the story events and generate changes in the narrative \cite{schjerlund2018design,gomezzara2019}. Many VR applications enable linear narratives that offer a structured and guided experience. The creators have full control over the storytelling and design of the sequence of events that users must follow. In contrast, interactive narratives empower users to make choices and influence the direction of the story, enhancing engagement and immersion \cite{green2014interactive,Riedl2013,scolari2009transmedia}. Temporal linear narratives are used mainly to guide users, including temporal instructions, topic animations, and session timers. The proposed VR system should support both kinds of narratives: it should enable linear narratives that bring ideas in a sequence, as well as interactive narratives that allow users to create, navigate, and edit topics.  

\paragraph{DG3: Displaying users' ideas and thoughts in VR to enhance reflection and creativity.}\label{sec: DG3} Since we aim to transform users' ideas and thoughts into a VR environment, it is important to leverage the immersive nature of the 3D space for reflection purposes. Employing a VR system with a deep spatial arrangement enhances users' sense of physical presence \cite{rosander2000interactivity, LindemanFidelity}, adding realism to the virtual environment. Most importantly, our proposed VR application should not overload users' virtual space in order to keep them focused and comfortable. Integrating visually engaging elements, such as dynamic animations and interactive functions, to represent users' provided information is a good way to elevate user engagement \cite{jerald2015vr}. Nonetheless, the information conveyed through these elements should be concise and lucid to ensure easy comprehension. Ensuring a clear and usable interface, the VR application will enable users to reflect on their ideas and boost their creativity. By leveraging the immersive capabilities of VR, we aim to create a more interactive and engaging way for users to interact with their thoughts. This approach should not only facilitate a deeper understanding of the content but also encourage users to explore and manipulate data in innovative ways. Visualizing ideas in VR should allow users to physically navigate through layers of information, possibly uncovering hidden patterns and connections that are not as apparent in traditional, flat data presentations. Ultimately, this could lead to more insights and solutions inspired by a novel way of experiencing and interacting with ideas and thoughts.


\paragraph{DG4: Enabling users to record, summarize, visualize, and expand relevant topics from their ideas in real-time.} Another main challenge of handling users' ideas is that the system needs to transcribe them and analyze them in real-time. For this purpose, we first need to employ robust speech recognition algorithms and natural language processing techniques to transcribe users' speech into accurate text data. Additionally, the system must implement summarization algorithms capable of extracting key themes and topics from the transcribed text while it is generated. Furthermore, the system should be able to recommend and propose ways to expand and connect the ideas emerging from users' speech. Lastly, the visualization of these topics must be intuitive and visually appealing, facilitating easy comprehension and interaction for users, as addressed in section \ref{sec: DG3}. 

\subsection{Example Scenario using VIVRA}\label{sec:example}
In this subsection, we provide an example that illustrates the practical functionality of the VIVRA system. Nicole, an undergraduate college student, wants to practice a presentation for her next human-computer interaction class. She has some ideas in mind and wants to brainstorm more of them, record the presentation, and then analyze the main topics covered to improve the presentation. Instead of writing her presentation down or recording herself to watch the video recording later, she uses VIVRA to record, summarize, visualize, and expand her ideas in the VR environment.

After Nicole launches VIVRA on her VR device, the application greets her by showing a start screen where she accesses the main scene. Once inside, she is immersed in a comfortable room with warm decorations, such as flowers, transparent windows, and wood doors. Nicole can see her animated hands and perform actions in the VR setting, including walking, turning, and teleporting. The VIVRA app shows the operation instructions on one of the walls of the room, showing Nicole how to use her voice, gaze, and controllers to interact with the VR environment. The app instructs Nicole to verbalize her ideas and thoughts so that they can be recorded. While speaking, she notices that after finishing a few sentences, a balloon with the mentioned topic comes out of a portal on the ground. Nicole can continue talking and see more balloons emerging from the portal or start exploring the detected topics. By examining the topics, Nicole can track the mentioned topics, think of the mentioned ideas, and reflect on some missed or new ideas that she can develop.

Once Nicole stops talking, she organizes the balloons by grabbing them and relocating them with her controller's grip button. After putting them in her preferred location, she starts analyzing the balloons' content and shapes. By looking at the size of the balloons, Nicole discovers that the balloon with the topic 'Technical Details' is too large, demonstrating that she talked too much about it. She clicks on that balloon with the trigger button on her controller and then sees the options to delete, view, and add more content. Nicole clicks on the view button and sees all the original transcripts referring to the technical details mentioned during her presentation. By using an LLM, the system also suggests ways to reduce this topic or break it down into other ideas. She realizes that the presentation included redundant information about the methods she was using. The LLM also suggests missing topics that could be relevant to her presentation by adding proposed balloons. Nicole uses the balloons to reflect on and improve her presentation. By checking every balloon in the room, Nicole has a rough version of her presentation in mind. 

In another run, Nicole also tries the VIVRA's recording mode to record her speech first and then analyze the transcript. Using this mode, she starts presenting a newer version of the presentation. After clicking start, VIVRA records her thoughts and ideas. Once she stops the recording, a balloon with the text 'Play' appears on her right. VIVRA starts playing Nicole's recording, and the interactive balloons with the presentation's topics emerge from the portals according to the moment these topics are mentioned. Nicole examines her presentation by listening to her recording and looking at the balloons. First, she discovers that some topics were not detected as she did not develop them enough. Second, she noticed that some moments of her presentation were too dense because four balloons were created at similar times. To see the balloons in a more organized view, Nicole first clicks a button on her controller to automatically arrange all balloons within her camera's view based on their height. As the balloons tend to float higher during the transcript playback, she examines the relative timing of the topics she discusses during her speech. After a deeper analysis, she realizes that she might have addressed the results too early and delved into background research too late in her presentation. The system also proposes and gives suggestions of how the ideas can be split and structured. She uses the same approach as she used during brainstorming to analyze and improve her speech.

By iterating through the entire process, Nicole finds that using VIVRA has greatly improved the preparation of her presentation. The immersive VR environment provided her with a unique way to brainstorm, organize, practice, and analyze her speech, ultimately leading to a more polished and effective presentation.

\subsection{System Workflow}\label{sec:work flow}
\begin{figure*}[!htb]
    \centering
    \includegraphics[width=\textwidth, trim = 0cm 2.4cm 0cm 2.4cm, clip]{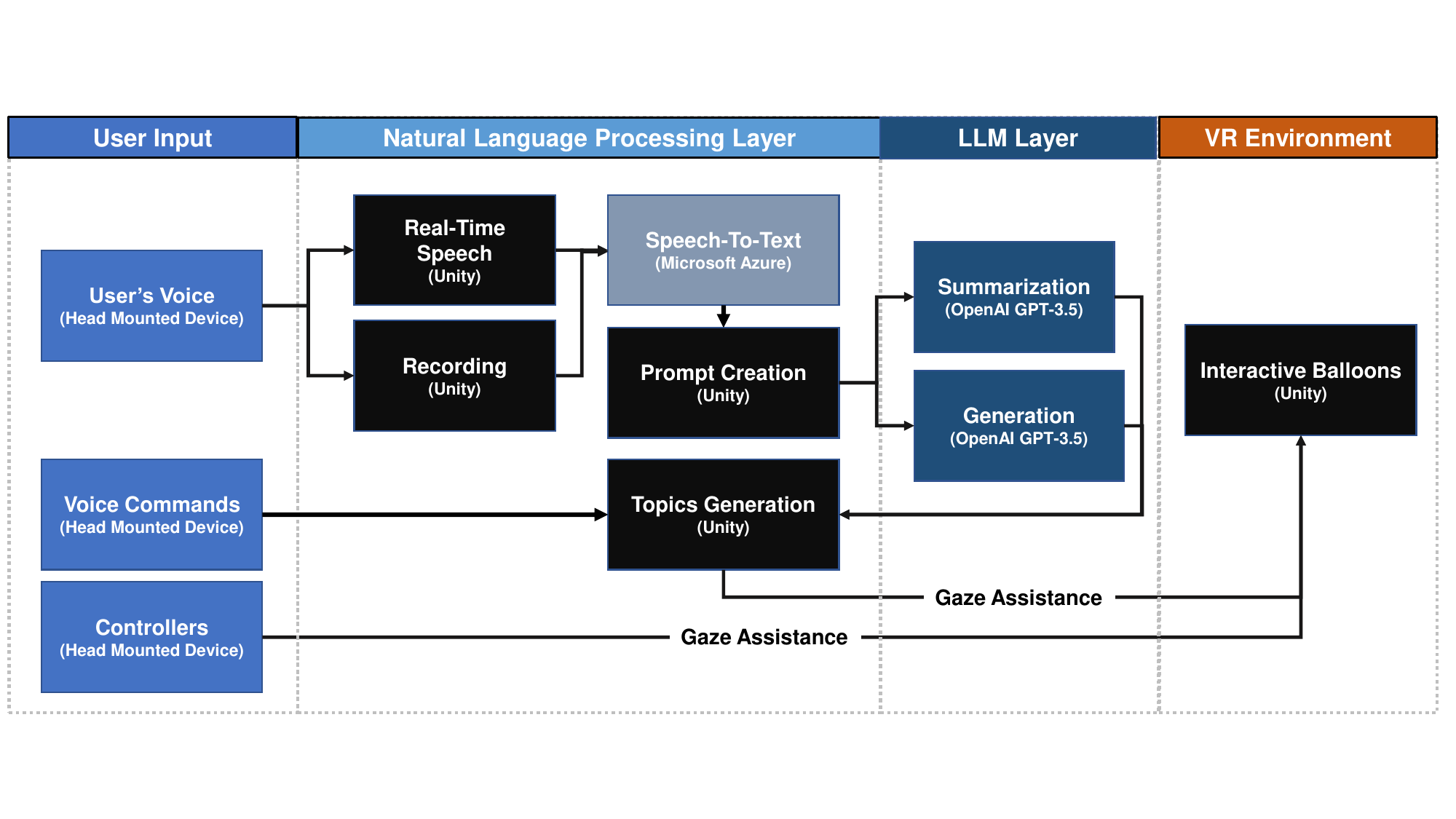}
    \caption{The workflow of VIVRA}
    \Description[The workflow of VIVRA]{The workflow of VIVRA}
    \label{fig:workflow}
\end{figure*}

Figure \ref{fig:workflow} and Algorithm \ref{algorithm:topic-generation} illustrates the workflow of VIVRA. The application tracks three possible user input entries. First, it will listen to users' voices to transcribe them and generate the topics. VIVRA will transcribe the users' speech using a speech-to-text conversion service. Once a sentence is finished, the system creates prompts using the user's speech transcription for the LLM. The goal is to extract and summarize the main topics mentioned by the user or the recorded speech. VIVRA then retrieves the key topics using the LLM and the original sentences associated with the topics mentioned in the transcripts. Second, VIVRA will also track users' gaze to render the ballons in their desired location. Third, the users can use the controllers to manually edit the balloons generated by VIVRA. 

\subsection{Key Features}\label{sec: key features}
\subsubsection{Multimodal Interaction with the balloons} \label{sec:balloonInteractive}
In this VR application, we used balloons as the main interactive objects. The balloons represent topics detected in users' speech. We decided to use balloons because they could represent encapsulated topics that float in the air. Alternative designs could include other shapes, such as stars, clouds, or birds.  Users can view the topic's transcripts by clicking the balloons and using the controllers. We designed the balloons with a degree of transparency ($\alpha=0.4$) so that users can see other balloons or objects behind them. Moreover, we added physical properties to the balloons so that when they collide with each other, they move slightly to the side, enabling clear visibility and avoiding overlaps. As mentioned in \textbf{DG1}, the goal of VIVRA is to enable multimodal interaction techniques, including voice, gaze, and controllers, to enable interactions with transcript data. We elaborate on their details in the following paragraphs:

\begin{algorithm}[!htb]
\caption{Workflow for Speech Processing and Balloon Generation in VIVRA}
\label{algorithm:topic-generation}
\begin{algorithmic}

\State \textbf{Input:} \( S_{\text{input}} \) \Comment{User's Speech}
\State \textbf{Output:} \( B_{\text{VR}} \) \Comment{Generated Balloons in the VR Environment}

\State \textbf{Terms:}
\State \( T_{\text{transcript}} \): Transcribed Text from User's Speech
\State \( P(T_{\text{transcript}}) \): Prompts generated from the transcribed speech
\State \( K = \{k_1, k_2, \dots, k_n\} \): Set of key topics extracted from the speech
\State \( R(k_i) = \{r_{i_1}, r_{i_2}, \dots, r_{i_p}\} \): Set of extracted original sentences related to each topic \( k_i \)
\State \( B(k_i) \): Balloon generated for topic \( k_i \). $B(k_i)$ includes the information of $k_i$ and $K_{\text{existing}}(k_i)$
\State \( R_{\text{existing}(k_i)} = \{R_{i_1}, R_{i_2}, \dots, R_{i_q}\} \): Set of all existing sets of sentences that have already been processed.
\State \( K_{\text{existing}} \): Set of all existing topics that have already been processed
\\

\State \textbf{Step 1: Speech Transcription}
\State Transcribe \( S_{\text{input}} \) using a Speech-to-Text service, obtaining \( T_{\text{transcript}} \).

\State \textbf{Step 2: Topic Extraction}
\State Generate prompts \( P(T_{\text{transcript}}) \) from the transcribed speech.
\State Use ChatGPT to extract key topics \( K = \{k_1, k_2, \dots, k_n\} \).
\State Retrieve the original sentences \( R(k_i) = \{r_{i_1}, r_{i_2}, \dots, r_{i_p}\} \) associated with each \( k_i \in K \).

\State \textbf{Step 3: Balloon Generation}
\For{each \( k_i \in K \)}
    \If{\( k_i \notin K_{\text{existing}} \)} 
        \State Create a new balloon \( B(k_i) \).
        \State Populate \( B(k_i) \) with \( R(k_i) \).
        \State Append \( k_i \) to \( K_{\text{existing}} \).
        \State Set $ R_{\text{existing}(k_i)} =  R(k_i)$
    \Else
        \State Identify the existing balloon \( B(k_{i}) \) related to \( k_i \).
        \State Append \( R(k_i) \) to \( R_{\text{existing}(k_i)}\).
    \EndIf
\EndFor

\State \textbf{Step 4: Display Balloons}
\State Create or update \( B_{\text{VR}} = \{B(k_1), B(k_2), \dots, B(k_n)\} \) in the VR environment at the designated locations. 

\end{algorithmic}
\end{algorithm}

\paragraph{Controller interactions} When the user's VR controller hovers over the balloon with the pointer, which is displayed as a graphic raycaster, a setting button appears on the top of the balloon. By clicking on the balloon, four interactive buttons will appear: ``View,'' in which users can view the topic's transcripts; ``Delete,'' which will delete the topic; ``Add,'' which can attach more transcripts by speaking; and ``Finish,'' in which users can exit the setting mode. Figure \ref{fig: balloon} (a) illustrates these interactive functions included in the balloon. Users can add transcripts to the balloons and extend their original transcripts. The new transcripts will be attached to the selected topic. After a new transcript is added, the balloon will also grow in size to reflect the changes. VIVRA allows users to manually operate the balloons to organize the space as they desire. They are able to grab the balloons by pressing the VR controller's grip button to relocate the balloons to their preferred position.

\paragraph{Voice-based balloon creation, deletion, and merging} As described in Section \ref{sec:work flow}, VIVRA enables the users' voice input method to create balloons. It provides a voice command function to help users add, edit, or delete the generated topics. The system will listen to users and wait for specific instructions designed for assistance. If the user announces an instruction, VIVRA will execute the specific instruction without sending it to the LLMs. The current system provides the following voice commands:
    \begin{itemize}
        \item \textit{Create a topic:} By saying "Create A," VIVRA would create a new balloon with the topic A attached to it and the note "You created this balloon by voice command." hidden in it. If the balloon already exists, the system will inform the user that the topic is already created. 
        \item \textit{Modify a topic:} When the system generates an incorrect or inaccurate title for the topic, users can change the title by saying "Change A into B." Then, the system will update the topic inserted in the balloon. The user can also say ''Expand A,'' so the LLM would create new ideas emerging from the topic A.
        \item \textit{Delete a topic:} Users can delete the topic A by calling the command "Delete A."  
        \item \textit{Merge topic:} It is possible that the system generates multiple topics from a single sentence or introduces a new topic even when there is a similar existing one. It could add topics that are highly related (e.g., ``Color'' and ``Paint'') or introduce a new topic from an existing one with a different spelling (e.g., ``Color,'' ``Colors,'' ``Coloring''). The user can merge two topics into one to avoid removing one topic and its attached transcripts. By saying "Merge A into B," topic A would be deleted and its transcripts would be attached to topic B's balloon. 
    \end{itemize}

\paragraph{Balloon generation and organization based on users' gaze} To ensure that the balloons are visible within users' view, we integrated the gaze function into VIVRA to acquire users' eye direction within the VR environment. With the assistance of gaze data, users are able to create balloons only within their eyesight. To prevent collisions, the balloons are generated in a random location within the users' field of view inside the room. Moreover, VIVRA plays a pop sound when the balloon is created to notify users that a new balloon is in the room.

\begin{figure}[!htb]
    \centering
    \includegraphics[width=1\textwidth]{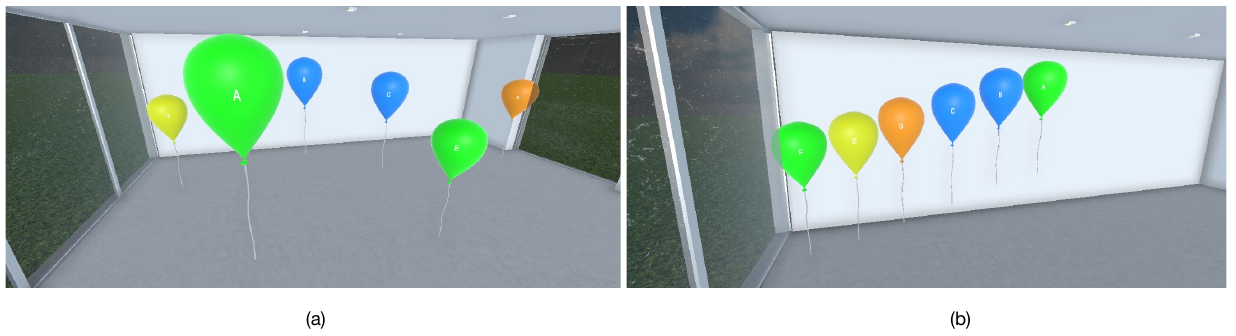}
    \caption{The one-click organization function using gaze: (a) before organization, (b) after organization.}
    \Description[The one-click organization function using gaze: (a) before organization, (b) after organization.]{The one-click organization function using gaze: (a) before organization, (b) after organization.}
    \label{fig:org}
\end{figure}

Moreover, moving balloons one by one could be inconvenient. To address this issue, we provided a one-click organization function that relocates all the balloons in the room close to the wall based on the user's gaze direction. Figure \ref{fig:org} shows the balloons' layout before and after pressing the button to organize the space. Balloons will float from their original position to their new position simultaneously and be rearranged based on their height. If the wall space is not enough for all the balloons, they will be arranged near another wall in clockwise order.

\subsubsection{Environment Design and Organizational Hierarchy}
We opted to design VIVRA as a room-scale space. This decision aims to balance providing ample space for users to operate comfortably and not overwhelming users with excessive space, ensuring that users are not required to move around too often. We chose to integrate the LLM into the environment, allowing users to speak aloud, with identified topics appearing as interactive balloons. While many systems use chatbots or virtual agents to represent LLMs, we designed VIVRA as a `smart room' to enable users to focus on the balloons rather than on a virtual agent. To add some depth, the balloons will appear at some distance from the users but will not touch the wall of the room. Lastly, to provide a rich narrative supported by storytelling elements (\textbf{DG2}) and text visualization in an appealing way in VR (\textbf{DG3}), we designed specific properties for balloons and space: 

\begin{itemize}
    \item \textbf{Temporal height view:} To offer a temporal view of the balloons that supports storytelling elements (i.e., \textbf{DG2}), we associated the balloons' height with their time creation. All the balloons will be at the same fixed height when created. As time goes on during the session, the balloon will gradually float higher, allowing users to look and estimate when they were created. The higher the balloon, the earlier it was generated. We controlled the balloon's height range so that it would not be uncomfortable for users to look at it. 
    \item \textbf{Dynamic scale adjustment:} The balloons have the same default size when they are created. If users add more details to an existing topic while speaking, VIVRA will check if the topic already exists and associate existing topics with new transcripts. The VIVRA app will append the new transcripts to the corresponding balloon. The balloon's size will then adjust according to the number of words in the combined transcripts. The greater the content within the topic, the larger the balloon will expand. By observing the sizes of balloons, users can tell the amount of notes they have within each topic. In this implementation, the balloons grow linearly according to the number of words in the transcripts. And after 300 words in total for the topic, the balloons will stop growing. Other approaches, such as applying geometric or logarithmic distributions, can be implemented to limit the increase of the balloons' size after many new words are included.
\end{itemize}

\subsubsection{Support for both narratives} \label{sec:twoNarrative} 
To provide a rich narrative supported by storytelling elements (\textbf{DG2}), users can choose linear or interactive narrative modes when using VIVRA. Once the application starts, VIVRA sets the interactive narrative mode by default. In this case, users create and interact with the main storytelling element, the balloons, with their voices, which enables an interactive narrative because every situation is solely triggered by users' voices or controllers. The interactive narrative ensures optimal flexibility so that users can think aloud. 

Users can enter the linear storytelling mode by recording their speech first and playing it later with the addition of interactive balloons. In this mode, users can only listen to the recorded speech and observe the balloons emerging from their portals, allowing users to have time to reflect on their speech. 

\subsubsection{Transition Animations} 
\begin{figure}[!htb]
    \centering
    \includegraphics[width=1\textwidth]{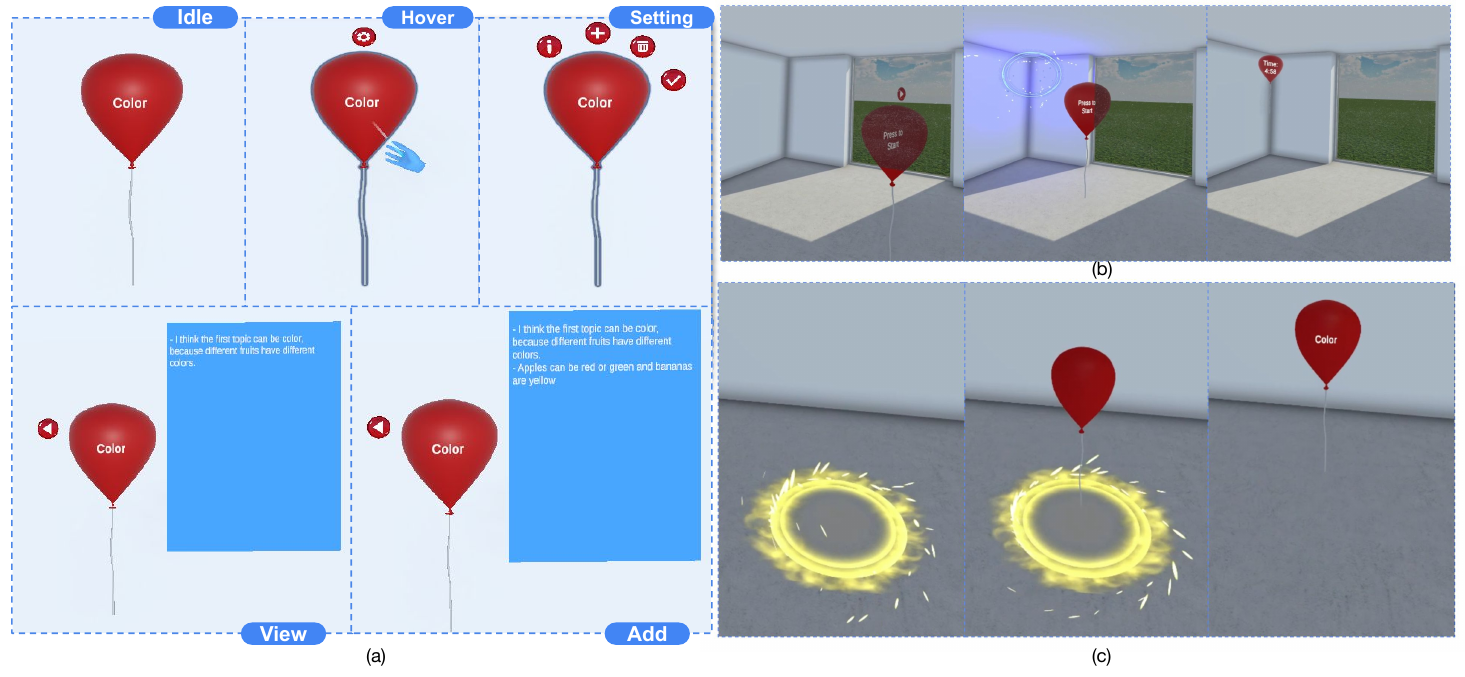}
    \caption{The functions and animation of the interactive balloon. (a) The functions of the interactive balloon. (b) The animation after a user clicks the Start balloon. The balloon will float to the corner and turn into a timer. (c) The animation of creating a balloon. A portal animation will emerge from the ground, and the balloon will ascend from the portal to the middle of the room.}
    \Description{The functions and animation of the interactive balloon. (a) The functions of the interactive balloon. (b) The animation after a user clicks the Start balloon. The balloon will float to the corner and turn into a timer. (c) The animation of creating a balloon. A portal animation will emerge from the ground, and the balloon will ascend from the portal to the middle of the room.}
    \label{fig: balloon}
\end{figure}

To support the linear narratives described in \textbf{DG2}, we implemented transition animations that provide an intuitive and simple way for users to understand the narrative flows. As shown in Figures \ref{fig: balloon} (b) and (c), we developed two main animations in VIVRA. The first set of transitions helps users navigate the initial instructions and create a new project. Once users start the app, they must click on a ``Start'' balloon located in the middle of the room. It will slowly float to the top corner of the room once clicked and will become a timer, helping users to manage and better organize their time. The second set of transitions provides animations for the interactive balloons. Once a new topic is created in the system, a portal will appear on the ground, and the balloon will ascend to the designated position (See Figure \ref{fig: balloon} (c)). After the balloon goes up, the portal will disappear.

\subsection{Implementation Details} 
We developed VIVRA in Unity 2021.3.27f1 and C\#. For user voice transcription, we utilized Microsoft Azure's standard-tier speech-to-text service. In detail, we employed the SpeechRecognizer class provided by the Microsoft Cognitive Services Unity API. By fine-tuning the SpeechRecognizer's recognized event, we processed every sentence from users' speech in real-time. We set the segmentation silence timeouts to 0.3s. We set all the other parameters to default. 

We used OpenAI's ChatGPT as the LLM for this implementation and integrated our system with OpenAI's GPT-3.5-turbo model. We used string variables to formulate the prompt and handle the outputs. After receiving the speech-to-text result from Microsoft Azure as a string, we inserted it along with the pre-existing topics into the prompt template we created, subsequently sending it to ChatGPT. We utilized local string arrays to store the detected topics and transcripts. Possible extensions could save these entities on databases and servers and use other large language models. After the system receives the output from ChatGPT, it parses the text string into two new string arrays, storing the topic and the transcripts. The system compares these newly detected topics with the existing topic array. If there is a new topic, the system will append it to the existing topic array. Otherwise, the system will attach the new transcripts to the corresponding ones. Algorithm \ref{algorithm:topic-generation} describes the pseudo-code of this implementation.

\section{Exploratory Evaluation}
We performed an exploratory evaluation to understand how users perceive and engage with the multimodal interaction techniques provided by VIVRA. Through the evaluation, our goal is to obtain participants' feedback on the interaction techniques. We identified advantages, obstacles, and potential improvements to implement in the system. 

\subsection{Participants}
We recruited 29 undergraduate students (11 female and 18 male) from a computer science class at a private university in the US. We compensated participants with extra credit for being part of the study. The sessions were conducted in November 2023. The study was reviewed and approved by our institution's IRB office (\# Anonymized).

\subsection{Procedure}
Each study session took approximately 30 minutes and was conducted in person in our laboratory. At the beginning of each session, the research assistants explained the purpose of the study and collected informed consent forms from the participants. Participants then completed a pre-treatment questionnaire that assessed their demographic information, creativity, and familiarity with VR and writing tools. Once the participants finished the survey, the research assistant guided each participant to a room with a Meta Oculus Quest 2 device connected to a computer that rendered the VIVRA application. The participants adjusted the headsets and started using the system with a brief tutorial explaining how to control the environment. 

In the VR environment, VIVRA showed the participants its key features and explained the interaction methods that they could use. Each participant took two minutes to explore the system using the instructions shown in the application. After familiarizing themselves with the interactions, they were asked to talk for three minutes about their plans for their coming vacation. While speaking, they were free to create, organize, delete, and edit topic balloons with their voice, controllers, and gaze. After the experiment was completed, the participants responded to a questionnaire and reported their experiences and opinions about the system. This final questionnaire included NASA's Task Load Index (TLX), which assesses workload on a 7-point scale \cite{hart1988development}, and the SUS usability \cite{brooke1996sus} scale that assesses a system's usability characteristics based on ten items using a 5-point scale. We also designed 5-point Likert scales to measure participants' satisfaction with the app, its effectiveness, its usefulness, and whether they will use it again. We also asked participants to respond to open-ended questions that addressed the advantages, obstacles, and suggestions for VIVRA.

\subsection{Results}
\subsubsection{Post-treatment questionnaire}
Results are presented in Table \ref{table:results}. Based on the NASA TLX items, the majority of the participants agreed that using VIVRA for this task was not physically demanding ($M=1.86, SD=0.88$) and not highly mentally demanding ($M=3.41, SD=1.13$). Moreover, most of the participants reported feeling successful in accomplishing the task ($M=4.59, SD=1.07$). Regarding VIVRA's usability, most participants found the system easy to use ($M=3.48, SD=1.13$) and did not need the support of a technical person ($M=3.55, SD=1.3$). The majority of the participants also agreed that it was helpful to see the main topics ($M=3.48, SD=1.10$) and summarize their thoughts when talking ($M=3.34, SD=1.06$). The majority also agreed that VIVRA's various functions were well integrated ($M=3.38, SD=0.89$). They did not find too much inconsistency ($M=3.62, SD=1.00$) or unnecessary complexity ($M=3.28, SD=1.11$) in the system. Regarding multimodal interactions with the balloons, participants found that adding balloons was a simple action with the combination of their voice and gaze ($M=3.41, SD=1.13$), and they also reported that it was easy to remove the balloons with their controllers ($M=3.14, SD=0.90$). 

However, participants reported lower scores when considering using this application again in the future. For example, many participants thought they did not get much from working with VIVRA ($M=2.79, SD=1.03$) or that they would have preferred to use a traditional tool instead ($M=2.24, SD=1.13$)\footnote{Reversed question}. During the experiment, participants reported that they did not utilize all their skills and abilities to complete this task and felt that they were not effective enough to work in this VR environment. Also, some participants reported that it was relatively difficult to edit balloons with their controllers.

\subsubsection{Open-ended questions}
Participants found VIVRA "very interactive" (P23) and "helpful just being able to talk and have your thoughts pop up" (P9). Some of them stated that it was nice "... to not have to use the traditional point-and-click system" (P9) and "not to need to type" (P28). P5 agreed that "... this method is far superior to offering the user a keyboard to type their ideas." Moreover, P14 stated "It was nice you are able just to speak your thoughts, and very little was demanded from you." P24 pointed out that the most significant advantage was "... being able to take notes based on my stream of consciousness." P25 found that "... most of the buttons used (edit, create, delete, etc.) were pretty standard, and I could instantly tell what they meant." Regarding VIVRA's VR spatial layout, P2 loved that he "... could look around and see what topics I had already spoken about." Also, "The idea of having balloons as well was really fun" (P27). P6 reported that "I can immerse myself in contemplating tasks because the outside environment wouldn't distract me. Also, it creates a space just for the user, so that the user can feel safe to think out loud." P29 also noticed that "... the size of the balloons increased with the frequency of the topics."

In terms of support for reflective thinking, VIVRA was "fun" (P17), and "... it made me go on and on about what was in the head and just let it out" (P17). "I like this VR application for brainstorming, especially for a paper. It captured many of the ideas I was talking about in balloons, and I liked how it categorized what I was saying into a balloon with the main idea. I was impressed by its ability to record \textit{'Parla'} accurately, a bar in Boston I mentioned. The controls were also easy to understand" (P27). P16 also reported that "... you don't really need to plan out your thoughts or follow an organizational structure". Overall, participants found that VIVRA enabled ``quickly taking down thoughts'' (P19), ``tracking thoughts'' (P7), ``summarizing words'' (P29), ``thinking aloud'' (P10), ``reflecting on ideas'' (P13), and ``generating more ideas'' (P24).

Despite these advantages perceived by the participants, others reported that the balloons sometimes "... appear where I was standing and I would be inside of a bubble" (P4). P8 reported that "... one of the balloons got so big that it made it hard to see the balloons behind." For P5, "I got very overwhelmed when using the balloons. Honestly, I would be talking, and out of nowhere, three balloons of different colors would appear at the same time, and I would completely lose track of my train of thought."

\subsection{Discussion}
Overall, the participants' feedback showed the success and potential of VIVRA and its novel multimodal interaction techniques. To solve the problems faced by the participants, we improved the physical properties of the balloons (Section \ref{sec:balloonInteractive}) and fixed the maximal size of the balloons (Section \ref{sec:balloonInteractive}) so that it would not block other balloons. To address the challenge some participants faced working with the interactive balloons while talking in real-time, we introduced the option to first record the speech and then play back the audio and interact later (Section \ref{sec:twoNarrative}). This approach not only prevents users from feeling overwhelmed but also ensures the delivery of linear narratives to meet their needs.

\section{Formal Evaluation}
After implementing improvements to VIVRA, we conducted a laboratory user study with ten participants in March 2023 to fully evaluate the system. To assess the effectiveness of the proposed visualization with the balloons, we ran this study incorporating other visualization techniques. We evaluated and compared the usability, effectiveness, and usefulness of these different visualizations. The study was reviewed and approved by our institution's IRB office (\# Anonymized). This second study aims to answer the following questions:

\begin{itemize}
    \item \textbf{RQ1:} Can users reflect more on their ideas or presentations by using the VIVRA visualization? 
    \item \textbf{RQ2:} How useful is VIVRA in facilitating the understanding and revision of users' ideas?
    \item \textbf{RQ3:} What advantages and challenges do users perceive when using VIVRA to revise their ideas?
\end{itemize}

\subsection{Participants}
We recruited ten participants from a private university in the US. The participants were all graduate students from the computer science department. All participants were fluent in English. In terms of gender, nine were male students and one was a female student. Participants volunteered for this study and were compensated with a \$15 gift card for their time. Participants completed an initial questionnaire to assess their perceived creativity skills and familiarity with technologies. Regarding their experience with VR, only two participants had used VR applications before. Most participants were confident in their creativity skills ($M=3.50, SD=0.92$). When asked about the challenges they frequently face when planning ideas, six mentioned ``generating ideas,'' five mentioned ``organizing ideas,'' and only one mentioned ``identifying ideas they want to share.'' We also asked them how familiar they were with creativity support tools. Four participants mentioned being familiar with mind maps, six with outline sketches, seven with whiteboards, and only one with AI tools. 

\subsection{Study Design}
Each study session took approximately 50 minutes and was conducted in person in our laboratory. We used a within-subject design in which all participants experienced the three conditions. We utilize a counterbalance design to minimize order effects and potential biases. Once they started the experiment, the conditions were displayed in random order.  

\begin{figure}[!htb]
    \centering
    \includegraphics[width=1\textwidth]{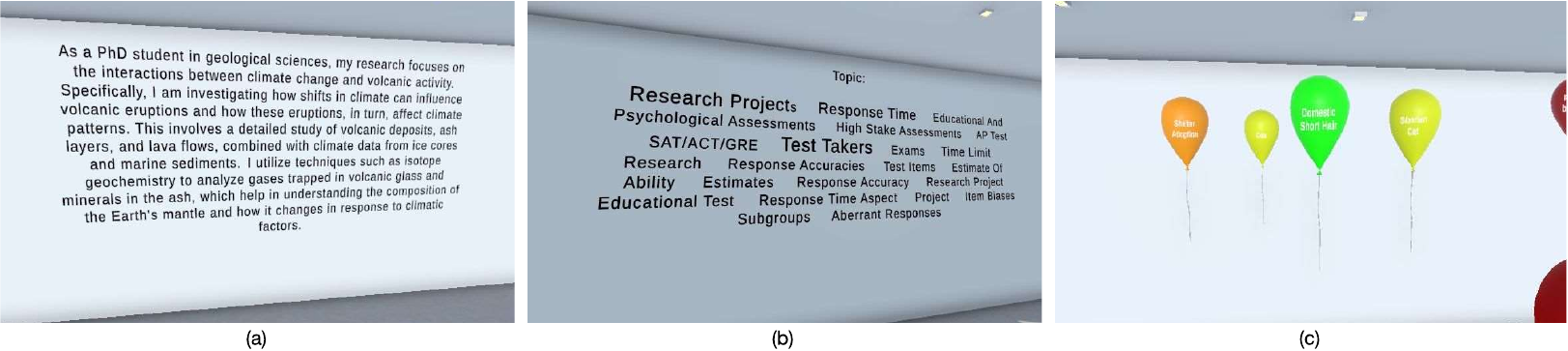}
    \caption{The three conditions of our evaluation: (a) Transcript, (b) Word Cloud, (c) VIVRA.}
    \Description[The three conditions of our evaluation: (a) Transcript, (b) Word Cloud, (c) VIVRA.]{The three conditions of our evaluation: (a) Transcript, which shows the entire transcript on a wall, (b) Word Cloud, which shows the most frequent keywords of the transcript, and (c) VIVRA, which shows the proposed interface with balloons.}
    \label{fig: conditions}
\end{figure}

\subsubsection{Conditions and Tasks} 
To empirically demonstrate the advantages of the proposed design in VIVRA, we tested the following three conditions that altered how the recordings were displayed in the VR application, as shown in Figure \ref{fig: conditions}:

\begin{itemize}
    \item \textbf{Transcript (baseline)}: The user will see her/his recording transcribed on a wall while listening to the recording. This condition represented a basic display of a recording.
    \item \textbf{Word Cloud}: The user will see her/his recording summarized in the form of a word cloud on a wall while listening to the recording. The words will appear, and the size will change based on the same rules described before. The same LLM implementation will be integrated into VIVRA to summarize the text.
    \item \textbf{VIVRA}: A full version of VIVRA that features balloons and voice commands. 
\end{itemize}

In each condition, the participants responded to one of the following three questions: (a) ``Tell me about your research project. What is it about?,'' (b) ``What was one big challenge during your first year as a graduate student?,'' and (c) ``What would be your dream research project? Please describe it and provide details.'' Each question was randomly assigned to one condition only.

\subsubsection{Procedure}
We followed the same routine as the exploratory evaluation to explain the purpose of the study and collect informed consent forms from participants. Participants completed the pre-treatment questionnaire to assess their demographics, perceived creativity skills, and familiarity with technologies. Once they completed the survey, the research assistant guided each participant to the experiment room and helped them put on a Meta Oculus Quest 2 headset. VIVRA showed the key features and explained each task to perform. The system asked the participant to respond to each one of the three task prompts and gave eight minutes for each. The system recorded the participants' voices while they answered each one of the prompts. The participants could record their responses again if they did not feel confident or made any mistakes. Each response exercise was independent of each other, so responses were not carried over to subsequent exercises. 

After the participants completed the recordings, VIVRA displayed each recording using the three experimental conditions. Participants listened to their recordings while viewing the visualization to be tested. Once the experiment was completed, participants responded to a post-treatment questionnaire that assessed their experience with each visualization. We kept a screenshot of each condition in the survey so that the participants could recall the interface correctly. We also incorporated a manipulation question to check that the participants distinguished the three visualization exercises.

As in the previous study, we included questions to measure the usability, enjoyment, advantages, and challenges of each visualization. We also asked the participants about the effectiveness of each visualization in achieving the design goals. We created a scale ``visualization assessment'' for users to judge how effectively these visualizations achieved the design goals (e.g., ``This visualization was effective in integrating both linear and interactive storytelling elements''). We calculated the Cronbach's $\alpha$ coefficients of these scales and found high consistency among the items ($\alpha_{Transcript} = .88, \alpha_{Word Cloud}=.95, \alpha_{VIVRA}=.92$). We also created a 5-point Likert scale with seven items to measure the usefulness of each visualization (e.g., ``Compared to other traditional methods, I would prefer to revisit recordings using this visualization.''). The Cronbach's $\alpha$ score on this scale showed reliable consistency ($\alpha_{Transcript} = .71, \alpha_{Word Cloud}= .65, \alpha_{VIVRA}= .64$). Moreover, we created seven 5-point Likert items to assess how the interface was better at supporting reflection activities than traditional applications (e.g., "think more ideas," "be more creative"). This scale scored high consistency ($\alpha_{Transcript} = .93, \alpha_{Word Cloud}=.92, \alpha_{VIVRA}=.95$). Lastly, we included open-ended questions to ask for more details of participants' experiences, feedback on the application, and the potential benefits and challenges of each visualization.

\subsection{Results}

\begin{figure*}[!htb]
\centering
    \begin{subfigure}[b]{.48\textwidth}
        \includegraphics[width=\textwidth]{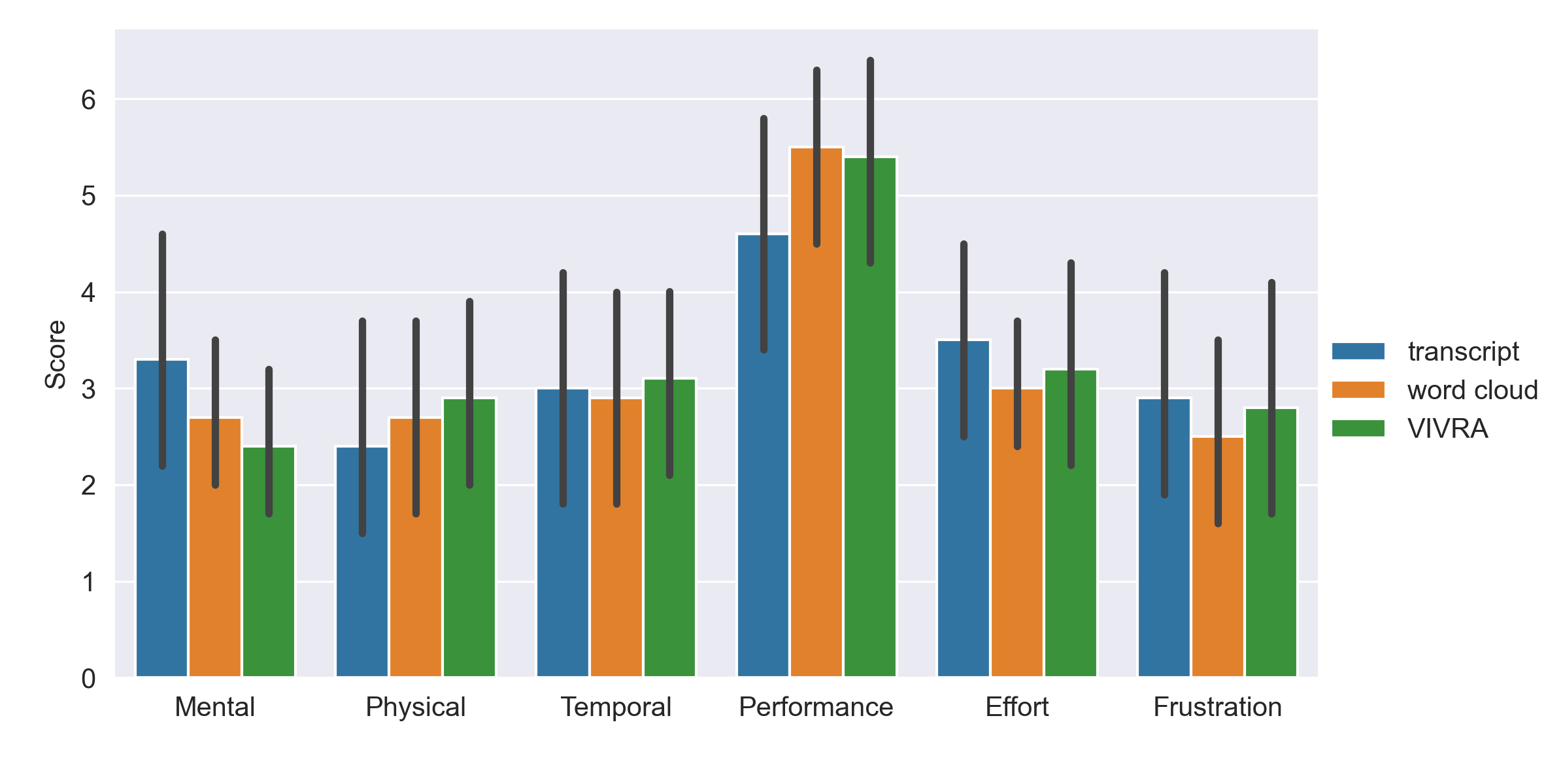}
        \caption{NASA TLX Results}
        \label{chart:result1}
        \Description[The post-treatment questionnaire result of NASA TLX.]{The post-treatment questionnaire result of NASA TLX.}
    \end{subfigure}\qquad
    \begin{subfigure}[b]{.47\textwidth}
        \includegraphics[width=\textwidth]{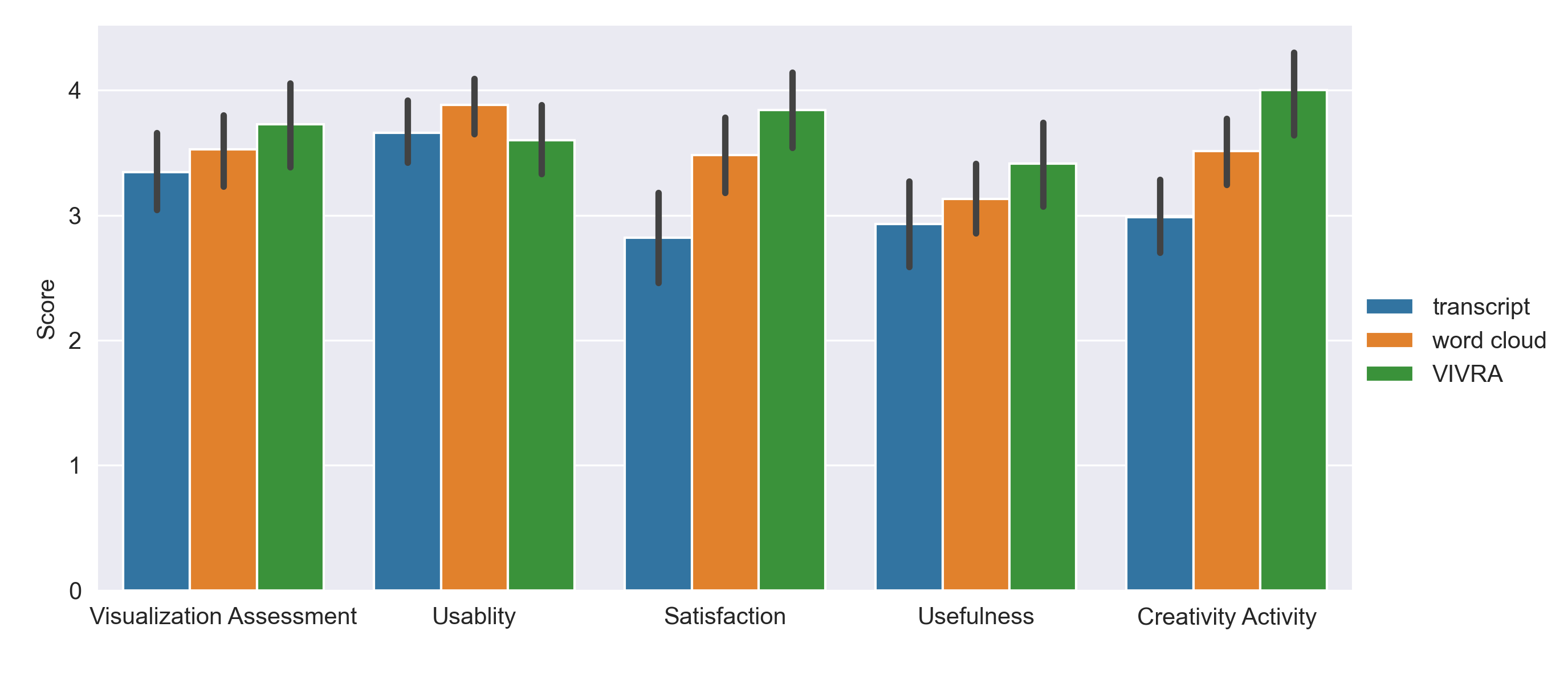}
        \caption{Other assessments }
        \label{chart:result2}
        \Description[Results of the post-treatment questionnaire]{Results of the post-treatment questionnaire}
    \end{subfigure}
\caption{The post-treatment questionnaire results}
\label{fig:post-treatment-results}
\end{figure*}

\subsubsection{Post-treatment questionnaire}
Fig \ref{chart:result1} and Fig \ref{chart:result2} shows the statistical results from the post-treatment questionnaire. When examining the NASA TLX items, we found that the VIVRA interface performed better in mental, effort, performance, and frustration levels than the transcript. Although the VIVRA visualization did not outperform the Word Cloud visualization in these categories, VIVRA required low mental efforts ($M=2.4, SD=1.26$) and physical demand ($M=2.9, SD=1.72$) and reached high performance ($M=5.4, SD=1.83$).

In other categories, the results indicated that the VIVRA visualization scored higher than the Transcript and Word cloud visualizations in visualization assessment, satisfaction, usefulness, and reflection activities. In general, the participants enjoyed visualizing their ideas more using VIVRA ($M=4.00, SD=1.18$) than with the Transcript ($M=2.70, SD=1.19$) and the Word cloud ($M=3.20, SD=0.98$) visualizations. Also, participants agreed that the VIVRA visualization ($M=3.70, SD=0.90$) was much better for annotating ideas than traditional methods, followed by the Word cloud ($M=3.10, SD=1.04$) and the Transcript ($M=2.30, SD=1.27$) visualizations. Regarding the questions that evaluated the design goals, participants agreed that the VIVRA visualization was effective in facilitating intuitive multimodal interaction ($M=3.90, SD=1.38$), transforming their ideas into the VR environment ($M=4.20, SD=1.17$), and integrating linear and interactive narratives ($M=3.80, SD=1.47$). The results of the questions from the reflection activities show that the participants found the VIVRA visualization excellent for tracking their ideas ($M=4.40, SD=1.20$), reflecting on their ideas ($M=4.30, SD=1.19$), and thinking of more ideas ($M=3.90, SD=1.30$). Although in the usability category, the VIVRA visualization ($M=3.60, SD=1.38$) did not score higher than the Transcript ($M=3.66, SD=1.34$) or Word cloud ($M=3.88, SD=1.14$) visualizations, the participants still considered the former to be easy to learn ($M=3.40, SD=1.50$). A potential explanation for the scores is that the VIVRA visualization required users to learn additional features and operations. 

To check whether these differences were statistically significant, we performed a one-way ANOVA with repeated measures to compare the effects of the three systems. The findings revealed a significant main effect on the score of reflection activities ($F=3.88, p<.05$) in which the VIVRA visualization scored the highest. We found no significant differences among the scores in all other categories. 

\begin{table}[!htb]
\centering
\small
\caption{Mean and Standard Deviation for Each Category and Subcategory}
\begin{tabular}{@{}cc@{}}
\vtop{\hbox{\strut}
\hbox{
\begin{minipage}[t]{0.5\textwidth}
\centering
\begin{tabular}{lccc}
\toprule
\textbf{Category} & \textbf{Condition} & \textbf{Mean} & \textbf{Std. Dev.} \\
\midrule
\multirow{3}{*}{Mental Demand} & Transcript & 3.30 & 2.06 \\
                            & Word Cloud & 2.70 & 1.34 \\
                            & VIVRA & 2.40 &  1.26\\
\midrule
\multirow{3}{*}{Physical Demand} & Transcript & 2.40 &  1.84\\
                            & Word Cloud & 2.70 & 1.70 \\
                            & VIVRA & 2.90 & 1.73 \\
\midrule
\multirow{3}{*}{Temporal Demand} & Transcript & 3.00 & 2.05 \\
                            & Word Cloud & 2.90 &  1.85\\
                            & VIVRA & 3.10 & 1.73\\
\midrule
\multirow{3}{*}{Performance} & Transcript & 4.89 & 2.09 \\
                            & Word Cloud & 5.89 & 0.93 \\
                            & VIVRA & 5.40 & 1.84 \\
\midrule
\multirow{3}{*}{Effort}     & Transcript & 3.50 & 1.72 \\
                            & Word Cloud & 3.00 &  1.15 \\
                            & VIVRA & 3.20 &  1.81 \\
\midrule
\multirow{3}{*}{Frustration} & Transcript & 2.90 & 2.02 \\
                            & Word Cloud & 2.50 &  1.72 \\
                            & VIVRA & 2.80 & 2.10 \\
\bottomrule
\end{tabular}
\end{minipage}}}
&
\vtop{\hbox{\strut}
\hbox{
\begin{minipage}[t]{0.5\textwidth}
\centering
\begin{tabular}{lccc}
\toprule
\textbf{Category} & \textbf{Condition} & \textbf{Mean} & \textbf{Std. Dev.} \\
\midrule
\multirow{3}{*}{Visualization  Assessment} & Transcript & \(3.34\) & \(1.34\) \\
                            & Word Cloud & \(3.53\) & \(1.26\) \\
                            & VIVRA & \(3.73\) & \(1.45\) \\
\midrule
\multirow{3}{*}{Usability} & Transcript & \(3.66\) & \(1.34\) \\
                            & Word Cloud & \(3.88\) & \(1.14\) \\
                            & VIVRA & \(3.60\) & \(1.38\) \\
\midrule
\multirow{3}{*}{Satisfaction} & Transcript & \(2.82\) & \(1.34\) \\
                            & Word Cloud & \(3.48\) & \(1.04\) \\
                            & VIVRA & \(3.84\) & \(1.13\) \\
\midrule
\multirow{3}{*}{Usefulness} & Transcript & \(2.93\) & \(1.47\) \\
                            & Word Cloud & \(3.13\) & \(1.24\) \\
                            & VIVRA & \(3.41\) & \(1.39\) \\
\midrule
\multirow{3}{*}{Creativity Activities} & Transcript & \(2.99\) & \(1.27\) \\
                            & Word Cloud & \(3.51\) & \(1.16\) \\
                            & VIVRA & \(4.00\) & \(1.42\) \\
\bottomrule
\end{tabular}
\end{minipage}}}
\end{tabular}
\label{table:results}
\end{table}

\subsubsection{Post-treatment open-ended questions}
Following the questionnaire, we asked participants open-ended questions to collect their feedback and suggestions. Many participants reported that the VIVRA visualization was effective in facilitating intuitive multimodal interaction techniques. P1 stated that the balloons have an advantage over ``previous methods that I cannot interact with..." because existing methods, such as the word cloud, just "... provided keywords, which do not give help in showing logic between ideas or improving them." Moreover, P1 mentioned that he could "... delete the unnecessary balloons." P6 wrote: "The interactive process is very interesting. The way to render and record topics is very attractive." Other participants reported that "...this visualization ensures better user interaction" (P7) and "...the buttons in the systems are very easy to learn and clear" (P10). P5 stated that the whole process of using the VIVRA system was "... fun and not mentally boring. I enjoyed moving the balloons and removing them. And also, I like that I can reorganize based on the different topics identified."  

Participants reported that the Transcript and Word cloud visualizations lacked interaction. P4 reported that "... some words I said showed on the wall are wrong. It would be better to add a correction option". Some participants complained about the visual aesthetics of these two visualizations. For example, P6 mentioned that "... the word font is fixed and can not be adjusted by the user. Sometimes I may want all the words to get bigger or smaller." Lastly, P9 indicated that "I could not interact with this word cloud, which means that I could not reorganize them or revisit the original transcript it represented.".

In terms of visualization, most of the participants agreed that the VIVRA visualization effectively ensured an immersive and comprehensive environment. They also confirmed that this visualization was effective in integrating both linear and interactive narratives. Furthermore, most participants reported that transforming text data into a 3D environment was effective using the VIVRA interface. P7 wrote, "The visualization is 3D; users can view the topic from a different angle; I think this is an important feature to generate novel ideas." P3 appreciated "... the function of expanding the balloons to a larger idea (with short phrases explaining each idea)." P9 added that the balloons appeared "...with an order, which helped me to reorganize my narratives and to see a logical structure between them."


Regarding reflection activities, most participants agreed that using VIVRA allowed them to keep track of their ideas. P1 mentioned that the "... balloons come with keywords that give me good ideas of what I said."  Most participants found the system helpful in reflecting, re-evaluating, and, in general, thinking more about their ideas. For instance, P9 confirmed that the balloons "helped me to reorganize my narratives and to see a logical structure between them." Last but not least, most participants thought that VIVRA could enable them to create more ideas. P7 mentioned that VIVRA can help users "... organize the topic they like and generate new ideas."

\section{Discussion}
In this study, we present a novel VR application designed to summarize, expand, and visualize users' ideas and topics using an LLM. By summarizing, transforming, connecting, and rendering users' ideas into digital objects, VIVRA encourages its users to reflect on their ideas, thoughts, and words. The ultimate goal is to support their creativity as a result of this reflection process. In this section, we delve deeper into the implications of this study, critically analyzing how VIVRA’s design contributes to enhancing interaction, creativity, and 3D visualization.

\paragraph{Enabling multimodal interaction to engage with ideas}
One of our design goals for VIVRA was to facilitate simple and intuitive multimodal interactions with users' ideas. Among the multiple input methods for VR systems, voice input is considered to be one of the fastest \cite{bowman2002text}. Moreover, the findings show that talking was an easy way for users to interact with their ideas and thoughts in the VR environment. Another intuitive interaction approach enabled by this VR application was the user's gaze, which was employed to control the locations where the balloons appeared. With the gaze-assisted voice interaction method integrated into VIVRA, users could simply speak and interact with the environment in a natural and straightforward way. For our interactive narrative mode, users were able to operate the system without controllers and organize the environment with voice commands in real-time, which is especially beneficial for users without VR experience. 

\paragraph{Supporting reflection in creativity}
We found that VIVRA offered a variety of methods to assist users in reflective thinking. Users could quickly grasp the main points of their ideas and thoughts using this application in a clear and appealing way. As a result, VIVRA allowed users to think more critically about their ideas. Furthermore, multimodal interaction methods offered users flexibility and usability to organize their ideas in the creative process, resulting in users more engaged with their ideas and deeper insights throughout the creative process. Integrating LLMs into VIVRA supported the system's capabilities to remember and understand users' ideas as well as their context. Our findings demonstrated that LLMs are capable of capturing topics from users' speech accurately and quickly and providing new connections and perspectives of their ideas. Lastly, the several physical properties that we designed for the balloons enabled users to sense some of the metrics of their speech, such as how much they have talked about a topic or when they first talked about a topic. Our findings demonstrate that VIVRA can improve the users' reflection process and support creative ideation processes.   

\paragraph{Users engaged with visualizing their ideas in VR environments}
Compared to 2D systems, VR environments can provide immersive and engaging experiences \cite{bowman2002text}, allowing users to interact with their ideas in a 3D space. The particular design of VIVRA fosters users' creativity and comprehension by integrating data with immersive interactions. Additionally, VIVRA offered our participants a sense of presence and embodiment, making it easier for users to explore and manipulate their thoughts in a natural and intuitive manner. The results suggest that the multimodal interactions provided by VIVRA improved the user experience and made the visualization process more dynamic and interactive. In particular, the participants of the second study reported that the VIVRA visualization provided a more enjoyable interaction and creative experience compared to the transcript and word cloud visualizations, which were essentially 2D text representations. Improving and better controlling the physical properties of VIVRA allowed for a significant improvement in the second study. By controlling the size and adding a physical body, the balloon visualization became visually clearer and easier to understand.


\subsection{Design Implications}
With its novel multimodal interaction methods and visualization experience, VIVRA has the potential for multiple applications. Our example scenario described how VIVRA can be used for educational purposes. In addition to practicing presentations, VIVRA can also be used to facilitate interactive lectures, workshops, or training sessions. For instance, teachers can use VIVRA to prepare their course lectures and iterate the topics. Moreover, the recording function can be used by students to listen to teachers' lectures in an interactive way. VIVRA can help organize the teachers' lectures as topics, and students can engage with the lecture topics visually, enhancing their understanding and retention of complex concepts.

The features of VIVRA can also be used in group meetings. Instead of taking notes with a notebook or laptop, which might distract users from focusing on the meeting, VIVRA can serve as the role of a note-taker by visually capturing key points discussed during the meeting. By visually examining the topics and employing the LLM, meeting participants can easily grasp the content, generate new ideas, and identify connections or contrasts between them. The balloons can serve as visual reminders of important topics or decisions made, making it easier to follow up on tasks after the meeting.

Another potential user case of VIVRA is for text analysis. For example, market researchers can utilize the VR application to analyze consumer feedback, survey responses, and focus group discussions. The balloons could represent key insights, trends, and opinions shared by the participants, allowing researchers to identify patterns, recommend or suggest new ideas, and make data-driven decisions.

\section{Limitations and Future Work}
We acknowledge the most prominent technical constraints of the current VIVRA version. One of the main limitations of VIVRA was the increasing number of balloons. Having many balloons around was overwhelming for the users. Users might have lost track of some of the balloons and found it difficult to manage them. Some irrelevant topics might also have been detected by the LLMs, generating unnecessary balloons. Although these balloons can be manually deleted by users, adding unwanted elements to the VR environment creates more obstacles to the experience. To mitigate this issue, future versions of this application should implement more sophisticated and precise prompts for its LLM. Another limitation we faced was the potential discomfort associated with interacting with the system through voice. While the system relies on voice input to detect speech topics, some users may feel self-conscious or uncomfortable speaking aloud. To improve this, future versions of this application could offer an alternative function to submit text files to the system. VIVRA was tested with short recordings, no longer than ten minutes. As such, it will be important to test its efficacy for longer sessions and recordings.

Regarding the study design, there are several potential concerns with the validity of our findings. Our participants were computer science students, who could have been more excited or interested in a VR application. Expanding the demographics of the participants in future deployment studies, including individuals with different technological interests and skills, will also help assess the ecological validity of VIVRA.

We also identified potential changes that future versions can adopt. For example, the balloons' colors were randomly chosen and can potentially represent other semantic features, similar to height and scale. Another potential change is visually grouping the balloons based on semantic relationships, which can be done by the LLM. VIVRA could integrate visible edges connecting balloons like mind maps or grouping them by colors. The grouping process can be performed manually by users or completed automatically by LLMs. Also, to make it more visually engaging, future work should expand the form of balloons into other objects' shapes, which could be created by state-of-the-art generative AI techniques. Lastly, future versions could offer more advanced interactions with LLMs within the VR environment. For example, the LLM could suggest potential narratives, changes in their ideas, or add multimedia content to the balloons to provide a rich and immersive multimedia experience \cite{Mills2022,siegel1995more,gomezzara2019}.

\section{Conclusion}
We have introduced VIVRA, a VR application that leverages multimodal interaction and LLM's capacity to record, summarize, expand, and visualize topics from users' ideas and thoughts. This VR application enables users to interact with their speech using voice commands, gaze tracking, and controller devices. Moreover, we designed a concise and captivating visualization, with the identified ideas represented as floating balloons. Through an exploratory study and a formal evaluation, we demonstrated the capabilities of VIVRA to enhance reflection, allowing users to generate and think more of their ideas. This study exemplifies the potential of integrating advanced technology and user-centered design to develop pioneering tools that support creativity, allowing users to engage and reflect on their ideas, thoughts, and words.

\begin{acks}
  This publication resulted from research supported by the Alfred P. Sloan Foundation (Award 2024-22427) and Microsoft Research Accelerating Foundation Models Research Program. The funders had no role in study design, data collection and analysis, decision to publish, or preparation of the manuscript.
\end{acks}

\bibliographystyle{ACM-Reference-Format}
\bibliography{sample-base}

\end{document}